\documentclass[iop]{emulateapj}

\usepackage{amsmath}

\usepackage{draftcopy}

\def\viibf{}
\def\mybf{}

\def\msun{$M_{\odot}$}

\def\sqig{$\sim$}

\def\ctscm2s{cts\,cm$^{-2}$\,s$^{-1}$}

\def\Fermi{{\it Fermi}}
\def\INTEGRAL{{\it INTEGRAL}}

\def\Swift{{\it Swift}}

\def\CGRO{{\it CGRO}}

\newcommand {\Pspindot} {$\dot{P}_{spin}$}

\newcommand {\nudot} {$\dot{\nu}$}
\newcommand {\nudotnu} {$\dot{\nu}/\nu$}

\def\Porb{P$_{orb}$}

\def\src{4U\,1538$-$52}
\def\igr{IGR\, J16393$-$4643}

\begin{document}
\def\subtitle{}
\submitted{}
\accepted{October 21, 2020}
\journalinfo{}

\title{
Superorbital Modulation in the High-Mass X-ray Binary
4U 1538$-$52, and Possible Modulation in IGR J16393$-$4643
}

\author{Robin H.~D. Corbet\altaffilmark{1,2,3},
Joel B. Coley\altaffilmark{4,5},
Hans A. Krimm\altaffilmark{6},
Katja Pottschmidt\altaffilmark{1,5},
Paul Roche\altaffilmark{7}}

\altaffiltext{1}{University of Maryland, Baltimore
County, MD 21250, USA; corbet@umbc.edu}

\altaffiltext{2}
{CRESST/Mail Code 662, X-ray Astrophysics Laboratory,
NASA Goddard Space Flight Center, Greenbelt, MD 20771, USA}

\altaffiltext{3}
{Maryland Institute College of Art, 1300 W Mt Royal Ave, Baltimore, MD 21217, USA}

\altaffiltext{4}
{Department of Physics and Astronomy, Howard University, Washington, DC 20059, USA. 
}

\altaffiltext{5}
{CRESST/Astroparticle Physics Laboratory, 
Code 661 NASA Goddard Space Flight Center, Greenbelt Rd., MD 20771, USA}

\altaffiltext{6}
{National Science Foundation, 2415 Eisenhower Ave., Alexandria, VA 22314, USA}

\altaffiltext{7}
{School of Physics and Astronomy, Cardiff University, The Parade,
Cardiff CF24 3AA, UK}

\begin{abstract}
Hard X-ray observations with the {\it Neil Gehrels Swift Observatory} Burst Alert Telescope (BAT) reveal
superorbital modulation in the wind-accreting supergiant
high-mass X-ray binary (HMXB) \src\ at a period of 14.9130 $\pm$ 0.0026 days that is 
consistent with 
four times the 3.73 day orbital period.
These periods agree with a previously suggested
correlation between superorbital and orbital periods in similar HMXBs.
During the \sqig14 years of observations 
the superorbital modulation changes amplitude, and since \sqig MJD 57,650 
{\viibf
it was no longer detected in the power spectrum,
although a peak near the second harmonic of this was present for some time.
} 
Measurements of the spin period of the neutron star in
the system with the \Fermi\ {\viibf Gamma-ray Burst Monitor} 
show a long-term spin-down trend
which halted towards the end of the light
curve, suggesting a connection between \Pspindot\ and superorbital modulation,
as {\viibf proposed for 2S\,0114+650}.
{\viibf However, an earlier torque reversal from INTEGRAL observations
was not associated with superorbital modulation changes.}
{\viibf
B and V band photometry from
the {\viibf Las Cumbres Observatory} reveals orbital ellipsoidal photometric variability,
but no superorbital optical modulation.}
However the photometry was obtained when
the 14.9130 day {\viibf period was no longer detected in the BAT power spectrum.}
We revisit possible superorbital modulation in BAT observations of 
IGR J16393$-$4643 but
cannot conclusively determine whether this is present,
although is not persistent.
{\viibf
We consider superorbital modulation mechanisms, and suggest that the Corotating Interaction
Region model, with small deviations from orbital synchronization, appears promising.}

\end{abstract}
\keywords{stars: individual (4U 1538$-$52, 4U 1538$-$522, QV Nor, IGR J16393$-$4643) 
--- stars: neutron --- X-rays: stars}

\section{Introduction}
\subsection{Superorbital Periods in Wind-accretion HMXBs}
Superorbital periods (periods longer than the orbital period)
in high-mass X-ray binaries {\viibf (HMXBs)} have been known for some time \citep[e.g.][]{Kotze2012}. 
The superorbital periods in the earliest known systems such as Her X-1, SMC X-1, and LMC X-4
could be accounted for as being due to precession of the accretion disk, formed
because the mass donating primary star fills, or is close to filling, its Roche lobe \citep[e.g.][and references therein]{Townsend2020}.
Such modulation had not been expected for systems where accretion occurs from the wind
of the primary. In these systems the low angular momentum of the accreted material should not, for most sources, lead
to the formation of a persistent accretion disk, although in some cases a transient disk might
form \citep[e.g.][]{Taam1988,Taam1989,Jenke2012,Romano2015b,Xu2019}.

However, there are now a number of wind-accreting HMXBs which do indeed
show persistent superorbital modulation. This was first found for
2S\,0114+650 by \citet{Farrell2006,Farrell2008}.
Subsequently, superorbital periods were found for the wind-accreting HMXBs IGR J16493-4348,
4U 1909+07 (= X 1908+075), IGR J16418-4532, and IGR J16479-4514 \citep{Corbet2010,Corbet2013}. 
There is still no generally accepted model to explain superorbital
modulation in these systems, although several have been proposed.
Detailed studies of IGR J16493-4348 have been made by \citet{Pearlman2019}
and \citet{Coley2019} and candidate models are reviewed therein.

\subsection{The Wind-accretion HMXB \src\label{sect:source_intro}}

\src\ (occasionally referred to as ``4U 1538$-$522'') is a bright eclipsing high-mass X-ray binary containing a
neutron star accreting from the wind of QV Nor, a B0 Iab star.
It is a well-studied system in X-rays and at other wavelengths.
For a recent review of observations of this system see
\citet{Hemphill2019}.
The neutron star spins at a period of \sqig526 seconds
with long-term spin-rate changes \citep[][]{Rubin1997,Baykal2006,Malacaria2020}.
The system exhibits regular total eclipses with a period of 3.73 days.
Although there have been a number measurements of the orbital
parameters from orbital Doppler shifts in the pulsation period,
the eccentricity is still rather uncertain.
An eccentricity of \sqig 0.18 has been reported by \citet{Clark2000}
and \citet{Mukherjee2006}, although an eccentricity of 0.08 $\pm$ 0.05
was reported by \citet{Makishima1987},
and a circular orbit may also
be possible \citep[e.g.][]{Corbet1993,Baykal2006,Rawls2011}.
The orbital ephemeris was recently updated by
\citet{Hemphill2019} from eclipse timing, and these authors
found marginal evidence for a change in the orbital period.
In this paper we use this orbital ephemeris for definition of
orbital period and eclipse center.

Optical photometry of \src\ was carried out 
by \citet{Ilovaisky1979} and \citet{Pakull1983}.
Both groups reported ellipsoidal variations due
to tidal distortion of the primary star with two
maxima and minima per orbital cycle. 
However, in addition to the periodic modulation there was
also cycle-to-cycle variability.
\citet{Rawls2011} also report on BVI photometry of \src\
and derive a low mass for the neutron star in the system
(0.87$\pm$0.07 \msun\ for an eccentric orbit and
1.00 $\pm$ 0.10 \msun\ for a circular orbit)
from a fit to the orbital photometric modulation.

We report here the detection using data from the \Swift\ BAT 
of superorbital modulation in \src\ at a period of 14.9130 $\pm$ 0.0026 days,
which is consistent with being exactly four times the orbital period
of the system. 
Measurements of the pulse frequency of
\src\ obtained with the \Fermi\ {\viibf Gamma-Ray Burst Monitor (GBM)} cover a large fraction of
the BAT light curve and we compare changes in superorbital modulation
with pulse frequency variations.
We also report on optical photometry of \src\
made with the Las Cumbres Observatory (LCO) telescope network.

\subsection{The Wind-accretion HMXB IGR J16393$-$4643}
{\viibf
IGR J16393$-$4643 is thought to be an HMXB containing a neutron star accreting
from the wind of its companion. For a summary of its properties
see \citet{Coley2015} and references therein. The spectral type of the
mass donor has not yet been determined, although \citet{Bodaghee2012} 
suggest that it may be a B-type main-sequence star.
The source has an orbital period of \sqig4.2 days as determined from
modulation of the X-ray light curve seen with the BAT \citep{Corbet2013,Coley2015}.
The minimum in the orbital modulation has been interpreted as an eclipse, and from measurements of this
\citet{Coley2015} found that the half-angle would be consistent with stars of
spectral type B0\,V and B0-5\,III.
However, \citet{Kabiraj2020} also investigated the X-ray
light curve of IGR J16393$-$4643 using \Swift\ X-ray Telescope observations and
proposed that the minimum in the orbital light curve, rather than being
caused by an eclipse, is due to absorption in a stellar corona.
We previously noted a possible superorbital period in
IGR J16393$-$4643
of 14.971 $\pm$ 0.005 days
on the basis of small peaks seen at this period and its second harmonic
in the power spectrum of the BAT light curve \citep{Corbet2013}.
While the statistical significance of the superorbital modulation 
was very low, we noted that power spectra of light curves from the {\it Rossi X-ray Timing Explorer} PCA
Galactic Plane Scan and \INTEGRAL\ IBIS observations
also both showed small peaks at the candidate superorbital period.
In addition, it was noted that the combination of superorbital and orbital
periods would be close to the 4.4 and 15.18 day orbital and superorbital
periods of 4U 1909+07. 
Here we reexamine the previously suggested superorbital period
in IGR J16393$-$4643 \citep{Corbet2013}. For IGR J16393$-$4643 we find that we cannot yet
definitely conclude whether superorbital modulation is 
present or not.
}

\section{Data and Analysis}

\subsection{BAT X-ray Observations of \src\ and \igr}

The \Swift\ BAT is a hard X-ray telescope that uses
a coded mask to provide a wide field of view \citep{Barthelmy2005}.
Here we use light curves from the \Swift\ BAT transient monitor \citep{Krimm2013}, which
are available shortly after observations have been performed and
cover the energy range 15 - 50\,keV. 
In this energy range, the Crab gives a count rate of 0.22 \ctscm2s.
The transient monitor light curves
are available with time resolutions of \Swift\ pointing durations 
(``orbital light curves''), and also daily averages.

For \src\ the observation durations of the orbital light curve
range from 64 to 2640 s and the mean duration is 666 s.
The light curve of \src\ considered here covers a time range
of MJD 53,416 to 58,880 (2005-02-15 to 2020-02-01).

{\viibf
For \igr\ the BAT light curve considered by \citet{Corbet2013}
covered the time range of  MJD  53,416 to 56,452  (2005 February 15 to 2013 June 9).
In contrast, the light curve of \igr\ now also covers until MJD 58,880, a duration of
5464 days, and hence is 2428 days (80 \%) longer. 
The durations of the individual observations of \igr\ again range from
64 to 2640\,s and the mean is 685\,s.
}

{\viibf Unless otherwise stated, our data analysis here uses the
orbital light curves rather than the daily averages.}
Our analysis was similar to that described in \citet{Corbet2013}.
We only use data for which the quality flag (``DATA\_FLAG'') was
0, indicating good quality. In addition, as in our previous work, we
removed points with very low fluxes which also had implausibly small uncertainties.
While changes to the transient light curve processing have reduced this problem,
some anomalous low flux points were still present before processing.

As in our previous work \citep[e.g.][]{Corbet2013, Corbet2017} we used discrete Fourier Transforms
to create power spectra {\viibf of} the BAT light curves. 
{\viibf Because the individual BAT observations vary considerably in exposure it is advantageous
to weight them when calculating the Fourier transform. 
\citet{Scargle1989} notes that the weighting of data points in calculating power spectra
can be compared to combining individual data points. The semi-weighted mean is a
generalization of the weighted mean \citep{Cochran1937,Cochran1954} which allows
for cases where the variation of data values is significant
compared to the size of their uncertainties. It is equivalent to the conventional weighted mean
when data value variation is small compared to the uncertainties.  
We therefore employ a technique based on the semi-weighted mean }
where each data point's contribution to the power spectrum is weighted by
a factor which depends on both the uncertainty on each point and the excess variability
of the light curve. Significance of peaks in power spectra are given as a false alarm
probability \citep[FAP;][]{Scargle1982} and period uncertainties are obtained via
the expression of \citet{Horne1986}. Although the calculation of the FAP depends
on the number of independent frequencies, which depends on the frequency resolution,
and this is not precisely defined for unevenly sampled data \citep[e.g.][]{Koen1990}, we have
found that using the inverse of the length of the light curve provides a good
approximation for BAT observations \citep{Corbet2017}.

\begin{figure}
\includegraphics[width=7.25cm,angle=0]{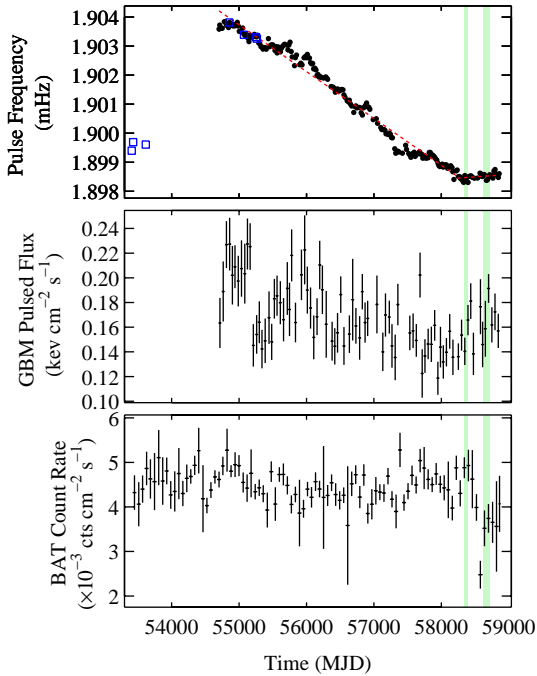}
\caption{Bottom: The \Swift\ BAT 15 - 50 keV light curve of \src. The light curve was taken from the
one day average light curves and then rebinned to a time resolution of 29.84 days
(eight orbital cycles, two superorbital cycles).
Middle: the pulsed 12 - 50 keV flux of \src\ as measured with the \Fermi\ GBM.
Fluxes are averaged over bins of length 44.76 days (12 orbital cycles, 3 superorbital cycles).
Top:
The pulse frequency for \src\ as measured with the \Fermi\ GBM. 
{\viibf The two dashed red lines
show the two linear fits to the pulse frequency given in Section \ref{sect:gbm_pulse}.}
{\viibf Pulse frequency measurements from \INTEGRAL\ observations by \citet{Hemphill2013}
are overplotted as blue boxes.} 
In all three panels the intervals during which LCO optical photometry was obtained are indicated
by the green boxes.
}
\label{fig:long_lc}
\end{figure}

\subsection{\Fermi\ GBM X-ray Observations of \src}

The \Fermi\ GBM \citep{Meegan2009} is a set of 12 sodium iodide and 2 bismuth germanate
scintillators primarily used to detect gamma-ray bursts.
In addition to this, the wide field of view and the high time resolution
of the GBM have been used to monitor bright X-ray sources, including
measurements of the spin periods of X-ray pulsars \citep[e.g.][]{Finger2009,Jenke2012,Malacaria2020}
in a similar way to previously undertaken with the {\it Compton Gamma-Ray Observatory} (\CGRO) 
Burst and Transient Source Experiment (BATSE).
We used measurements on the spin frequency of \src\ that are provided
online by the GBM
team\footnote{https://gammaray.msfc.nasa.gov/gbm/science/pulsars.html}.
These measurements cover from MJD 54,690 (2008-08-12) to 58,852 (2020-01-04).

\subsection{LCO Optical Photometry of \src}

The LCO \citep{Brown2013} is a global network of 0.4-m, 1-m, and 2-m telescopes
that operate as a single observatory. 
{\viibf
Observations of QV Nor, the optical counterpart of \src, 
were obtained in two groups. The first set of observations was obtained between 
MJD 58,345 to 58,390 (2018-08-15 to 2018-09-29) using 1-m telescopes, 
and the second set was obtained between 58,624 to 58,724 (2019-05-21 to 2019-08-29)
using 2-m telescopes. Observations were primarily obtained with B and V filters with a small
number of observations with other filters. Here we report only on the B and V observations.
During the first set of observations 24 B and 21 V measurements were obtained, with individual
exposures of 15s and 5s respectively. During the second set of observations, 63 B and V measurements
were obtained with individual exposures of 30 s and 15 s respectively.
A log of the observations is given in Table \ref{table:lco_observations} and
the times of the two sets of observations are indicated in Figure \ref{fig:long_lc}.
}
Photometric measurements were obtained using the ``X-ray Binary New Early
Warning System'' pipeline \citep[XB-NEWS;][]{Russell2019}. XB-NEWS provides 
photometry using several different aperture sizes. Here, based on advice from the XB-NEWS team,
we use the measurements obtained with apertures which were 1.0 times the Point Spread Function Full-Width
half-maximum (FAP\_1P0) for all stars on each frame, although this differs from frame to frame.
The first {\viibf set of observations}, which were taken with a 1m telescope and used {\viibf shorter exposures,} 
were found to have considerably larger error bars than the {\viibf second set of} observations which were obtained with a 2m telescope {\viibf with longer exposures.}
{\viibf For the 
first set of observations 
the mean uncertainties are 0.02 mag. for V and 0.046 mag. for B,
while for the
second set of observations 
the mean uncertainty of the V observations is 0.005  mag. and the mean uncertainty of the B
observations is 0.01 mag.}
There is also a shift between the mean B magnitude by about 0.3 between the two sets of
observations, although no such shift is seen for the V-band observations.
For these reasons we primarily use the data from the second set of observations that were obtained with
{\viibf longer exposures on} a 2m telescope.
{\viibf The 100 day time span covered by this set of observations corresponds to \sqig 26.8 orbital and \sqig6.7 superorbital cycles.}

\section{Results}
\subsection{BAT X-ray Results on \src}

\begin{figure}
\includegraphics[width=7.25cm,angle=270]{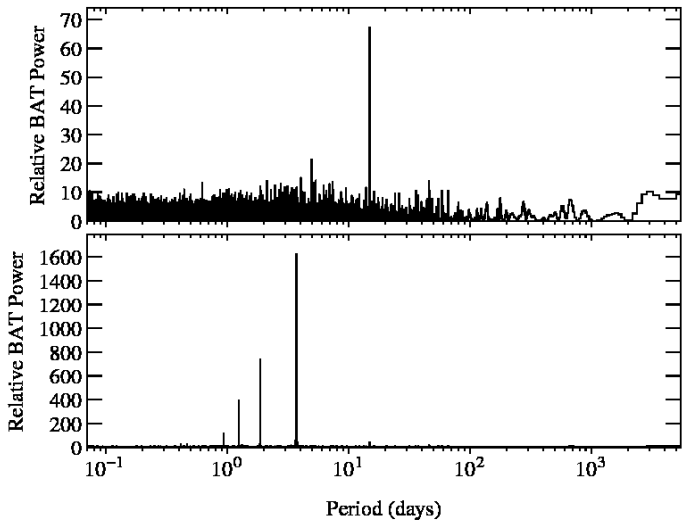}
\caption{Bottom: Power spectrum of the entire \Swift\ BAT light curve of \src.
Top: Power spectrum of the edited \Swift\ BAT light curve of \src.
Times around the eclipse in the system were removed before computation
of the power spectrum. 
{\viibf
For both panels the power is normalized
to the average power in the frequency range plotted.
}
}
\label{fig:bat_power}
\end{figure}

The BAT light curve of \src\ is shown in Figure \ref{fig:long_lc} (bottom panel).
The power spectrum of this, from a period of 0.07 days to
the length of the light curve (5464 days) is shown in Figure \ref{fig:bat_power} (bottom panel).
This shows a very strong peak at the orbital period of \src\ at 3.72836 $\pm$ 0.00006 days,
consistent with the period of 3.728354 $\pm$ {\viibf 0.000009} days reported by \citet{Hemphill2019}.
Large peaks are also seen at the 
second (1.86416 $\pm$ 0.00001 days $\Rightarrow$ 
\Porb\ = 3.72832 $\pm$ 0.00003 days), 
third (1.242777 $\pm$ 0.000007 days $\Rightarrow$
\Porb\ = 3.72833 $\pm$ 0.00002 days)
), 
and fourth (0.932084 $\pm$ 0.000006 days $\Rightarrow$
\Porb\ = 3.72834 $\pm$ 0.00003 days)
harmonics of the orbital period. 

In addition to the peaks associated with the orbital period, we also noted a much smaller peak
at 14.9126 $\pm$ 0.0031 days at a height of \sqig44 times the mean power level 
{\viibf over the entire period range} and
\sqig 24 times the local power level. % 10 - 30 days
While significantly smaller than the peaks related to the orbital period, 
the height of this peak compared to the local power level 
corresponds to a nominal FAP of \sqig3$\times 10^{-6}$ and so is highly statistically significant.
While some types of period search techniques can produce signals at
``sub-harmonics'' of the true modulation
period, i.e. multiples of the intrinsic modulation period, this effect
does not occur in direct Fourier-based analyses which quantify
the sine-wave components of the modulation.
Therefore, modulation at an integer 
{\viibf multiple}
of the orbital period cannot be an artifact
of the analysis procedure. However, the 3.73 day period of the system must be
the orbital period as this is derived from both the Doppler modulation of the
neutron star pulse period and the occurrence of total eclipses.

\begin{figure}
\includegraphics[width=7.25cm,angle=270]{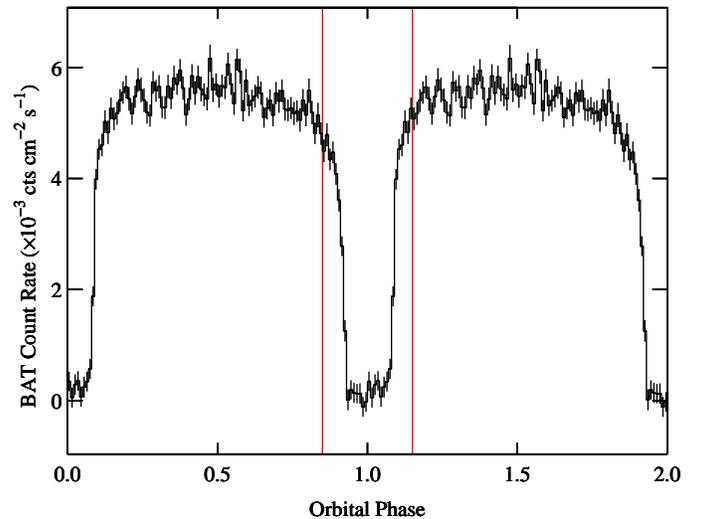}
\caption{\Swift\ BAT light curve of \src\ folded on the orbital period
of 3.728354 days, and phase 0 of MJD 57,612.401 \citep{Hemphill2019}.
The vertical red lines indicate the phase region around the eclipse that
was removed before computing the modified power spectrum.
}
\label{fig:bat_orbit_fold}
\end{figure}

To investigate this longer-period modulation we then wished to 
remove the potentially confounding effects of the orbital modulation on the light curve.
The BAT light curve folded on the orbital period is shown in Figure \ref{fig:bat_orbit_fold}.
The dominant contribution to orbital modulation is the total eclipse.
Therefore, to remove this modulation from the light curve we excluded observations
between orbital phases of 0.85 to {\viibf 1.15}, where phase {\viibf 1.0} is the center of the eclipse.
The power spectrum of this edited light curve for the same period range is shown in
the upper panel of Figure \ref{fig:bat_power}. The orbital modulation is no longer
visible in the power spectrum, and the strongest peak
is now at 14.9130 $\pm$ 0.0026 days and its height has increased to
\sqig67 times the mean power level.
Taking the local power level, the relative peak height is \sqig33 with an
associated FAP of \sqig4$\times 10^{-10}$.
The ratio between superorbital period and orbital period is thus
3.9999 $\pm$ 0.0007. i.e. the ratio is within 0.018\% of exactly
a factor of four.

As a check that the window function caused by the excision of data 
around the eclipse in the light was not causing apparent modulation
at four times the orbital period, we created a light curve with the
same sampling, but with the data values replaced by a pure sine wave
modulation with a period equal to that of the orbital period. 
For the sampling of the original light curve, the power spectrum
of the ``fake'' light curve only showed a peak at the orbital period.
For the sampling of the light curve with eclipse phases removed, the power
spectrum of this fake light curve did also show additional peaks at {\em higher}
frequency harmonics of the orbital period, but not at lower frequencies, including around
the \sqig14.9 day superorbital period.

\begin{figure}
\includegraphics[width=7.25cm,angle=0]{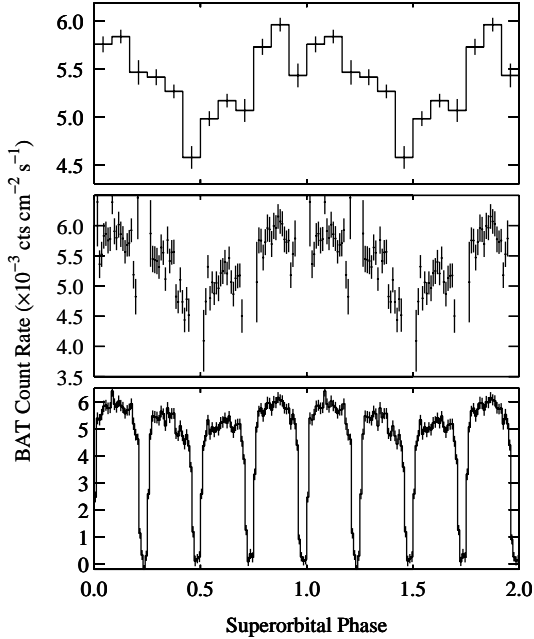}
\caption{\Swift\ BAT 15 - 50 keV light curve of \src\ folded on the superorbital period of 14.913 days.
Phase zero corresponds to MJD 56,106.4.
{\viibf Bottom: all data folded into 100 bins to illustrate the effects of the eclipse.
Middle: data with times of eclipses removed folded into 100 bins to show the gaps
produced by removing the eclipses.
Top: data with times of eclipses removed folded into 12 bins to make the overall modulation clearer.
}
}
\label{fig:bat_super_fold}
\end{figure}

To characterize the modulation near 14.9 days in \src\ we also
made a sine wave fit to the modified light curve.  
Leaving the period free we find a value of 14.9148 $\pm$ 0.0019 days,
consistent with the period obtained from the power spectrum.
The mean count rate
is 0.0055 \ctscm2s\ (approximately 25 mCrab) and the sine wave
semi-amplitude is 0.00046 $\pm$  0.00003 \ctscm2s, approximately 2.1 $\pm$ 0.1
mCrab. The epoch of maximum flux is MJD 56,106.4 $\pm$ 0.2, which is
0.25 $\pm$ 0.2 days after the time of eclipse center, i.e. an orbital phase
of 0.07 $\pm$ 0.05.
We note that because of
the exact ratio between the superorbital and orbital
periods and the relative phasing, we are {\em not} measuring the flux at either the implied
maximum or minimum of the {\viibf superorbital} modulation, if it is sinusoidal, because these would occur during the eclipses.
The BAT light curve folded on this superorbital period is shown
in Figure \ref{fig:bat_super_fold}. 
{\viibf
In the bottom panel we plot the unedited light curve folded into 100
bins and
the orbital eclipses are strongly present.
In the middle panel we plot the folded light curve with the eclipses
removed in which the times of the eclipses are seen as gaps.
In the top panel
we plot the light curve with eclipses removed folded into 12 bins to more clearly show the modulation on
the superorbital period.
}

\begin{figure}
\includegraphics[width=8.5cm,angle=0]{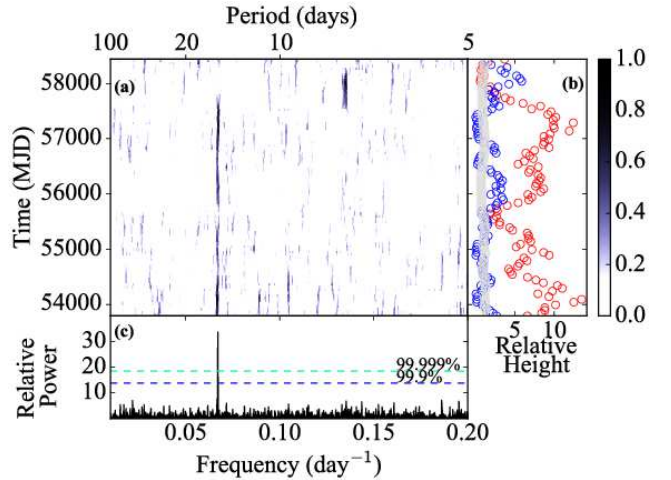}
\caption{
(a) Dynamic power spectrum of the BAT light curve of \src\ with orbital phase intervals
around the time of eclipse (Figure \ref{fig:bat_orbit_fold}) removed.
The superorbital period of 14.913 days corresponds to a frequency of 0.0671 days$^{-1}$.
(b) Red points: relative height of the peak at the superorbital period
{\viibf
to the mean power of values shown in panel (a).
}
Blue points:
relative height of the peak at the second harmonic of the superorbital period
{\viibf
to the mean power of values shown in panel (a).
}
Grey points, mean
power. All of these are relative to the mean power of all power spectra.
(c) Coherent power spectrum of the entire light curve. White noise significance levels (1 - FAP) are marked.
{\viibf
The power is normalized
to the average power of the coherent power spectrum in the frequency range plotted.
}
}
\label{fig:bat_2d_power}
\end{figure}

To investigate the long-term behavior of this modulation we calculated
a dynamic power spectrum. 
{\viibf
To obtain this we calculated the power spectrum of a segment of the light curve,
and then shifted the start time of the segment by an offset and calculated
the power spectrum of that. 
The individual power spectra are thus not statistically independent.
The choice of the length of the light curve segments is a compromise between
the segment being long enough to enable modulation to be significantly detected,
but not being so long that changes in the modulation become smeared out.
We experimented with different segment lengths and found that 
in order for the superorbital modulation to be seen it was necessary to
use segments at least several hundred days long.
Here we show results using a segment length of 750 days, although the overall
results are essentially the same if somewhat different segments lengths are used.
}
The resulting 
{\viibf dynamic power spectrum is}
shown in Figure \ref{fig:bat_2d_power}.
It is found that there are two main intervals when the superorbital modulation
is most prominent, from around the start of the light curve to approximately
MJD 55,000
{\viibf 
when the modulation is briefly not detectable in the dynamic power spectrum},
and then from approximately MJD 56,000 to 57,500
{\viibf when it is detected again at a comparable level to the earlier portion of
the light curve.}
Most recently
the superorbital modulation is not clearly visible in the dynamic power spectrum.
Instead, the maximum power occurs near the second harmonic of the superorbital period.
To characterize this we calculated power spectra of the most recent observations
with different start 
{\viibf and end}
times to determine what selection of time range maximized 
the signal at the harmonic.
From this, we find that the power near the harmonic has its strongest signal relative
to
the mean power if observations {\viibf from} approximately MJD 57,650 (2016-09-19)
{\viibf 
to 58,800 (2019-11-13) are used.
}
For this subset, the {\viibf strongest} peak in the power
spectrum is at a period of
{\viibf 
7.430 $\pm$ 0.006 days
}
and has a height of 
{\viibf 17.8}
compared to the mean power over the period
range from 0.07 days to 
{\viibf 1145 days,}
the length of that subset of the light curve,
and the associated FAP is 
{\viibf 3$\times10^{-4}$.}
This period is close to half the superorbital period derived from
the entire light curve of 7.4565 $\pm$ 0.0013 days, although they are
nominally statistically different from each other at \sqig {\viibf 4}$\sigma$.

\begin{figure}
\includegraphics[width=7.25cm,angle=0]{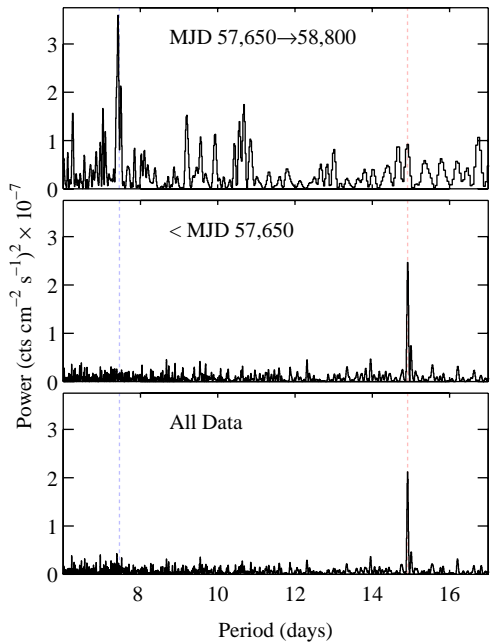}
\caption{
Power spectra of the BAT light curve of \src\ covering the region of
the superorbital period to the second harmonic of this for -  Bottom: the entire light curve.
Middle: the light curve from the start until MJD 57,650. Top: the
from MJD 57,650 until {\viibf 58,800}.
The vertical dashed red and blue lines show, respectively, the superorbital
period of 14.9130 days and its second harmonic at 7.4565 days.
Here, the power is plotted as absolute power, rather than as relative to the mean power level.
Times around the orbital eclipse were excluded.
}
\label{fig:bat_three_power}
\end{figure}

\begin{figure}
\includegraphics[width=7.25cm,angle=0]{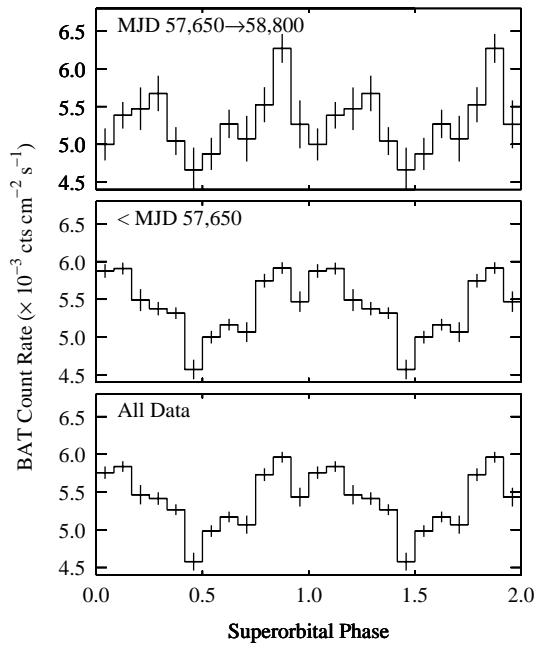}
\caption{
The BAT light curve of \src\ folded on the superorbital period of
14.9130 days for: Bottom: the entire light curve.
Middle: the light curve from the start until MJD 57,650. Top: the
light curve from MJD 57,650 until {\viibf 58,800}.
Times around the orbital eclipse were excluded.
}
\label{fig:bat_three_fold}
\end{figure}

To illustrate this change, in Figure \ref{fig:bat_three_power} we show
the power spectrum of the BAT light curve covering the region of
the orbital period to the second harmonic for (i) the entire light curve,
(ii) the light curve from the start until MJD 57,650, and (iii) the
power spectrum of the light curve from MJD 57,650 until {\viibf 58,800}.
From this, the superorbital period is only seen in the first
two power spectra, and the absolute power of this modulation is stronger 
for time interval ``(ii)''. 
In Figure \ref{fig:bat_three_fold} we show the BAT light curve of \src\
folded on the 14.9130 day superorbital period for the same three time ranges.
Thus, there appears to have been a change in the shape of the superorbital modulation
{\viibf during the interval MJD 57,650 to 58,800 
with the profile now having a more pronounced maximum near a superorbital phase of \sqig 0.85
and a hint of a secondary small maximum near phase \sqig 0.3.}

\subsection{Fermi GBM Pulse Frequency Measurements of \src\label{sect:gbm_pulse}}

The GBM measurements of the pulse frequency {\viibf \src}, together with the pulsed flux,
are shown in Figure \ref{fig:long_lc} (top and middle panels respectively).
We also overplot the pulse frequency measurements reported from \INTEGRAL\
measurements by \citet{Hemphill2013}.
Although there is considerable fluctuation during the GBM observations, the overall trend is
for an initial decrease in spin frequency (``spin down'') followed
by a flattening of the frequency changes and a slight trend towards
increasing spin frequency (``spin up''). Due to the fluctuations
and the uncertainties on the frequency measurements, the time of
the transition between the two trends is not well defined, but
appears to occur near approximately MJD 58,230 {\viibf (2018-04-22)}.
Fitting linear trends to the frequency measurements before and after this
time we find spin frequency changes of \nudotnu\ =  
-9.8 $\pm$ 0.1 $\times$10$^{-12}$ s$^{-1}$
and
+1.0 $\pm$ 0.5 $\times$10$^{-12}$ s$^{-1}$.
For comparison \citet{Rubin1997} found an overall spin-up trend of  \nudotnu\ \sqig +9.6 $\times$10$^{-12}$ s$^{-1}$ with BATSE
over a four year period, although with fluctuations around this trend, and an implied earlier spin-down rate of  \nudotnu\ = 
-8 $\times$10$^{-12}$ s$^{-1}$ from more scattered measurements made with a variety of satellites.
{\viibf Neither
the pulsed flux measured with the GBM nor the BAT flux measurements (Figure \ref{fig:long_lc}) show
any large change associated with the change in \nudot.}

{\viibf
The \INTEGRAL\ pulse frequency measurements from \citet{Hemphill2013}, the last few of which
overlap with the GBM measurements, show that a much more rapid spin-up occurred between
\sqig MJD 53,620 and 54,690 (Figure \ref{fig:long_lc}). This is also not associated with
any large change in the BAT flux.
}

\subsection{Results of LCO Optical Photometry of \src}

\begin{figure}
\includegraphics[width=7.25cm,angle=0]{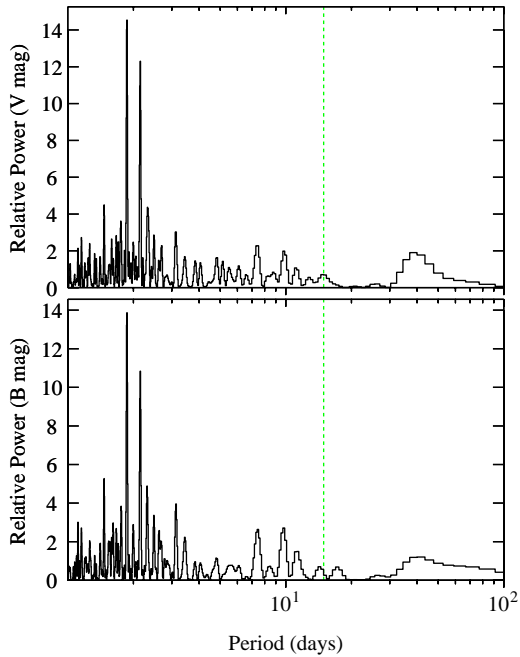}
\caption{
Power spectra of the LCO optical photometry of QV Nor, the optical
counterpart of \src. Only observations obtained with LCO 2m telescopes are included.
The strongest peak in both plots near 1.86 days is at half the orbital period, as expected
for ellipsoidal variability.
The dashed green lines indicate the superorbital period seen with the BAT.
}
\label{fig:lco_power}
\end{figure}

\begin{figure}
\includegraphics[width=7.25cm,angle=0]{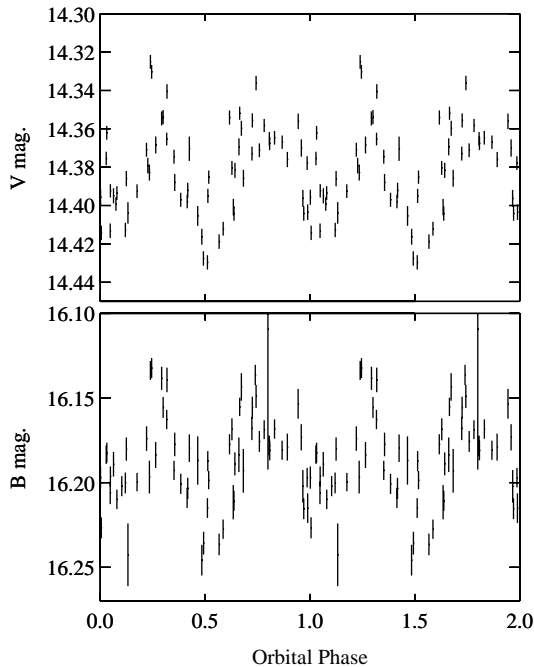}
\caption{
LCO optical photometry of QV Nor, the optical counterpart of \src, folded on the orbital period
of 3.728354 days. Phase zero corresponds to MJD 57,612.401.
B-band observations are shown in the bottom panel and V-band observations are shown in the top
panel. Only observations obtained with LCO 2m telescopes are included.
}
\label{fig:lco_fold}
\end{figure}

To search for modulation in the LCO photometry we calculated the power
spectra of the light curves, using only observations obtained with 2m telescopes 
{\viibf with longer exposures}
because of their
smaller uncertainties, and these are shown in Figure \ref{fig:lco_power}.
For both the B- and V-band light curves the strongest peak is at half the orbital
period. This is expected because the orbital modulation is driven by ellipsoidal
variability which gives two maxima and two minima per orbit. In Figure \ref{fig:lco_fold}
we show the B- and V-band light curves folded on the orbital period.
In the power spectra, however, we do not see any indication of modulation at
the superorbital period. 
For the B- and V-bands we determine 90\% upper limits on the semi-amplitude of
modulation on the superorbital period of 0.01 mag. for both wavebands.
In both power spectra we see a second peak 
near a period of 2.15 days. 
As a check on whether the secondary peak may be due to aliasing,
{\viibf
for each LCO photometric light curve (B and V) we fitted 
a sine wave with a period fixed to half
the orbital period, and then subtracted this from the
light curve.
} 
While the minimum at phase 0.5 due to ellipsoidal modulation is expected
to be deeper than the minimum at phase 0.0, 
{\viibf subtracting a single sine wave
with a period of half the orbital period was found to be sufficient to
remove }
the orbital modulation from {\viibf both} power spectra. 
In addition,
the power spectra no longer showed
modulation near 2.15 days, suggesting that it is indeed due to aliasing of the orbital
modulation. The semi-amplitudes of the fitted sine waves to the orbital modulation were 0.024 $\pm$ 0.003 mag.
and 0.021 $\pm$ 0.003 mag. for the B- and V-bands respectively.

We also calculated power spectra of {\viibf only} the observations obtained with {\viibf the} 1m {\viibf telescopes}, and these do
not show a significant detection of the orbital modulation on their own.
Calculating weighted power spectra of the combined 1m and 2m observations together did not
show significant differences from the power spectra from just the {\viibf 2m} observations.

\subsection{Results of BAT Observations of \igr}

\begin{figure}
\includegraphics[width=7.25cm,angle=270]{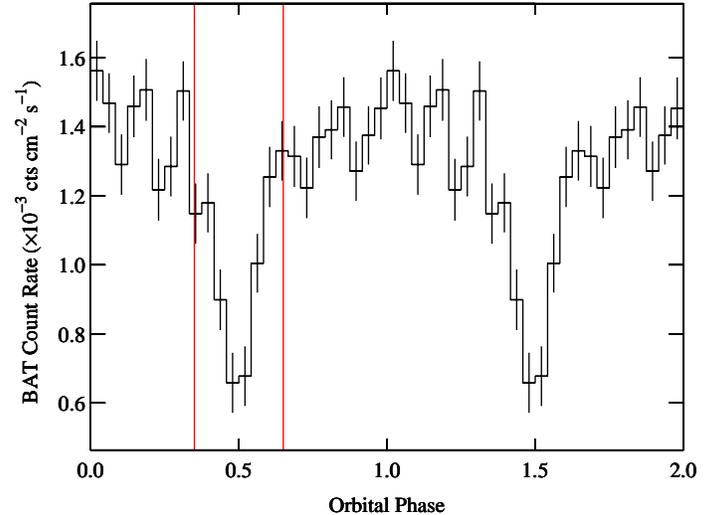}
\caption{\Swift\ BAT light curve of \igr\ folded on the orbital period of 4.2376 days.
Phase zero corresponds to MJD 54,352.50.
The vertical red lines indicate the phase region around the minimum that
was removed before computing the modified power spectrum.
}
\label{fig:16393_bat_orbit_fold}
\end{figure}

\begin{figure}
\includegraphics[width=7.25cm,angle=0]{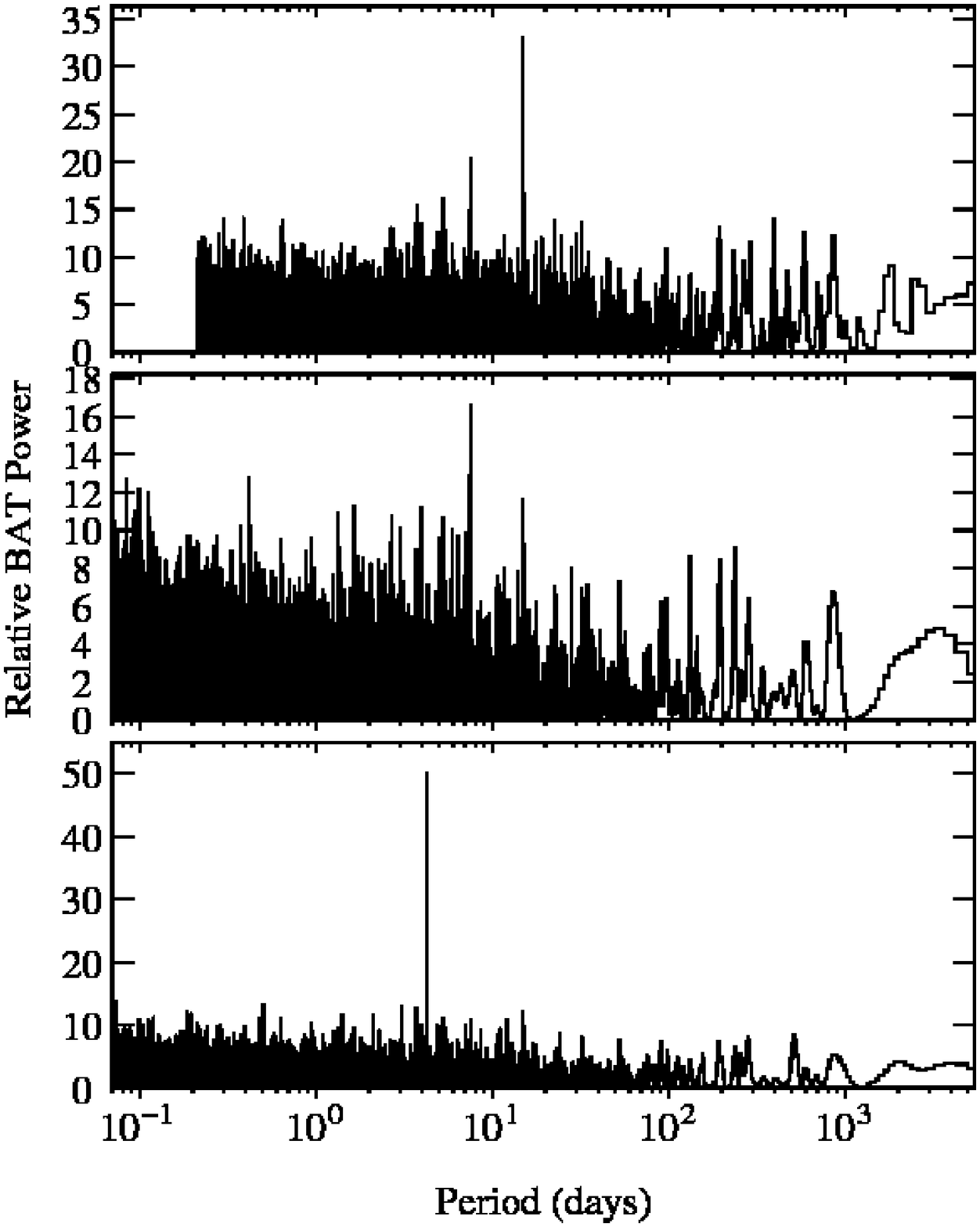}
\caption{Bottom: Power spectrum of the entire \Swift\ BAT light curve of \igr.
The strongest peak is at the orbital period of \sqig4.2 days.
Middle: Power spectrum of the edited \Swift\ BAT light curve of \igr.
Times around the eclipse in the system were removed before computation
of the power spectrum.
Top: Modified version of the middle plot. Each point in the power spectrum is replaced
with the sum of the original value plus the next two higher harmonics.
{\viibf
For all panels the power is normalized
to the average power in the frequency range plotted.
}
}
\label{fig:16393_bat_power}
\end{figure}

\begin{figure}
\includegraphics[width=7.25cm,angle=270]{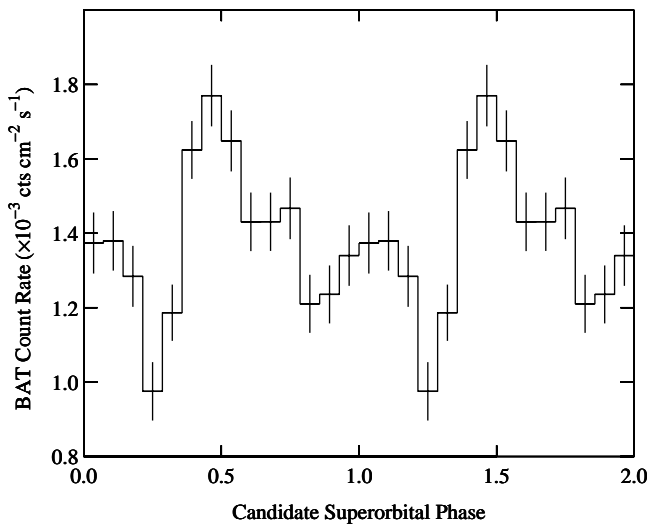}
\caption{\Swift\ BAT light curve of \igr\ with times around orbital
eclipse removed folded on the 
candidate superorbital period of 14.981 days. Phase zero corresponds to MJD 52,851.33. 
Data from times around the minimum of the orbital modulation
as marked in Figure \ref{fig:16393_bat_orbit_fold} were removed.
}
\label{fig:16393_bat_super_fold}
\end{figure}

{\viibf
In Figure \ref{fig:16393_bat_power} we show three versions of the
power spectrum of the BAT light curve of \igr. 
}
The lower panel shows the power
spectrum of the entire light curve. This shows a very prominent peak
at the orbital period of 4.2376 $\pm$ 0.0002 days, which is consistent with our
previously derived period and the more precise value of 4.23810 $\pm$ 0.00007 days
obtained by \citet{Coley2015} from eclipse timing.
However, no other strong peaks are seen in the power spectrum.
We then, in a similar fashion to our
analysis of \src, removed points around the minimum of the orbital light
curve. 
{\viibf To facilitate comparison with the results presented in \citet{Corbet2013} we adopt the same
convention for the definition of the orbital minimum and place this at phase 0.5 at a time of
MJD 54,352.50 and remove times corresponding to }
{\viibf orbital phases 0.35 to 0.65,} as illustrated in Figure \ref{fig:16393_bat_orbit_fold}. 
{\viibf We also employ the orbital period that we determine from the BAT power spectrum.}
The power
spectrum of 
{\viibf 
the light curve with times around orbital minimum}
is shown in the middle panel of Figure \ref{fig:16393_bat_power}.
The two highest peaks, although not very strong, are at the 
fundamental (14.9814 $\pm$ 0.0055 days)
and second harmonic
(7.4903 $\pm$ 0.011 days $\Rightarrow$ P$_{super}$ = 14.9805 $\pm$ 0.0022 days)
of the previously suggested superorbital period. We next, in order to perform a more
sensitive search for non-sinusoidal modulation, took the power spectrum of the light
curve with orbital minimum points removed, and replaced each data point 
{\viibf in the power spectrum}
with
the sum of the point and its second and third harmonics.
This is shown in the top panel of Figure \ref{fig:16393_bat_power}.
The highest peak is now at 14.980 days, near the previously proposed 14.971 $\pm$ 0.005 day superorbital period, with
a relative height of 33.
The light curve folded on this period is shown in Figure \ref{fig:16393_bat_super_fold}.
The possible superorbital modulation is similar to that determined previously 
\citep{Corbet2013}, and is rather ``bumpy'' and not sinusoidal, consistent with
the presence of multiple harmonically-related peaks in the power spectrum.

\begin{figure}
\includegraphics[width=8.5cm,angle=0]{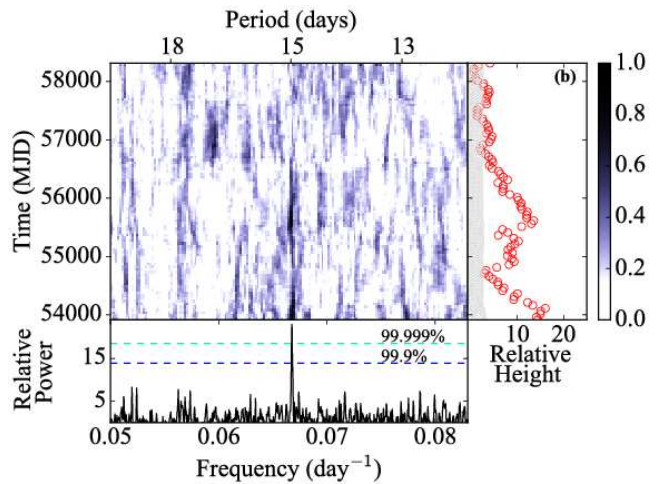}
\caption{
(a) Dynamic power spectrum of the BAT light curve of \igr\ with orbital phase intervals
around the time of minimum flux (Figure \ref{fig:16393_bat_orbit_fold}) removed.
Harmonic summing as employed in the upper panel of Figure \ref{fig:16393_bat_power}
was employed. The candidate superorbital period of 14.981 days corresponds to a frequency
of 0.067 days$^{-1}$.
(b) Red points: relative height of the peak at the harmonic sum of the superorbital period
compared to the mean of all power spectra. 
Grey points, mean
power of individual power spectra compared to the mean power of all power spectra.
(c) Coherent power spectrum of the entire light curve. White noise significance levels (1 - FAP) are marked.
{\viibf
The power is normalized to the mean power in the frequency range plotted.
}
}
\label{fig:16393_bat_2d_power}
\end{figure}

\begin{figure}
\includegraphics[width=7.25cm,angle=0]{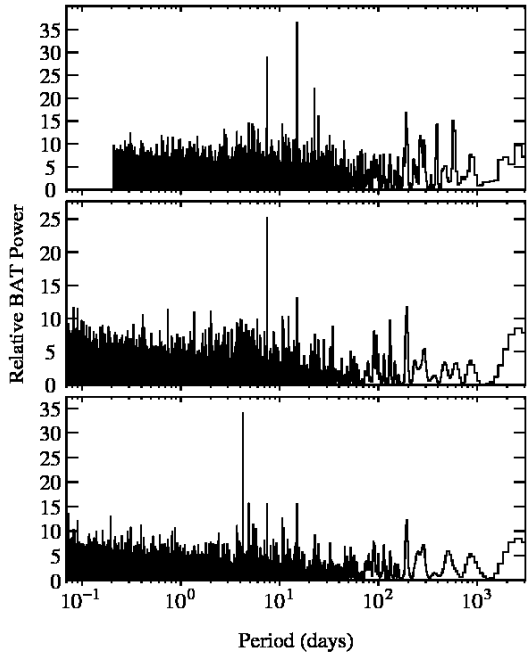}
\caption{Bottom: Power spectrum of the \Swift\ BAT light curve of \igr\ up to MJD 56,500.
Middle: Power spectrum of the edited \Swift\ BAT light curve of \igr\ up to MJD 56,500.
Times around the eclipse in the system were removed before computation
of the power spectrum.
Top: Modified version of the middle plot. Each point in the power spectrum is replaced
with the sum of the original value plus the next two higher harmonics.
}
\label{fig:excise_16393_bat_power}
\end{figure}

To investigate the long-term properties of this candidate signal, we calculated
the dynamic power spectrum of the BAT light curve of \igr\ in a similar way
to that employed for \src. However, in addition to removing data around the
orbital minimum of the light curve, we also applied harmonic summing up to the
third harmonic in the same
way as in the upper panel of Figure \ref{fig:16393_bat_power}. 
Light curve subsections of length 1000 days were used, with increments between sections
of 10 days.
The resulting
dynamic power spectrum is shown in Figure  \ref{fig:16393_bat_2d_power}.
The candidate period is only apparent in approximately the first half of the
light curve. 
{\viibf We therefore investigated the power spectrum of the BAT light curve of \igr\
using only data obtained before MJD 56,500 (2013-07-27). In a similar way to the power spectra
of the entire light curve shown in Figure \ref{fig:16393_bat_power}, we calculated
power spectra of the full light curve up to this time,
with orbital minima removed, and the summed harmonic power spectrum of this subset of
the light curve. The resulting power spectra are shown in Figure \ref{fig:excise_16393_bat_power}.
We find that there is only a very modest increase of the size of the peak at the candidate
superorbital period.
}

\section{Discussion}

{\viibf

\subsection{\src\label{sect:source_discuss}}

The BAT light curve of \src\ has a highly significant superorbital
modulation at a period consistent with exactly four times the orbital period.
In order to compare the superorbital properties of \src\ with
the previously known systems, in Figure \ref{fig:super_ratio} we plot superorbital period against
orbital period for wind-accretion supergiant HMXBs, and also the
ratio of superorbital period to orbital period as a function of orbital
period.
It was noted in \citet{Corbet2013} that, with only five data points,
there appeared to be a correlation between superorbital period
and orbital period. Although there are still only six sources, the
addition of the periods for \src\ is consistent with such a relationship.
It is notable that both the orbital period and superorbital period of \src,
and hence the ratio between the two, are very similar to those of IGR J16418-4532.
It is also striking that for \src, within the uncertainties, the superorbital
period is consistent with being exactly four times the orbital period 
(Figure \ref{fig:super_ratio}, Table \ref{table:sources}) with the
implied time of superorbital maximum and minimum for sinusoidal modulation occurring
during the orbital eclipses. 

\begin{figure}
\includegraphics[width=8.0cm,angle=0]{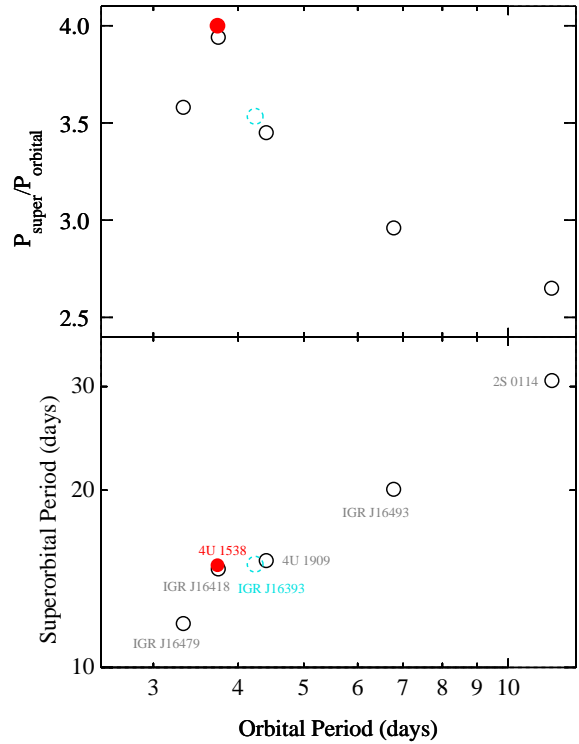}
\caption{
Bottom: superorbital periods of wind-accretion HMXBs plotted against orbital period.
{\viibf Truncated sources names are shown. Full names are listed
in Table 1. Parameter uncertainties are smaller than symbol size.}
Top: ratio of superorbital period to orbital period plotted against orbital period.
For both plots the parameters of \src\ are marked with a solid red circle.
The parameters of the candidate modulation in \igr\ are marked with a dashed
cyan circle.
}
\label{fig:super_ratio}
\end{figure}

The strength of the superorbital modulation in \src\ is found to vary,
but the length of the period and the phasing of the modulation 
do not show obvious changes during most of the observations. 
However, in the most recent observations (after \sqig MJD 57,650) there is a distinct change in the
modulation profile which results in the decrease of the peak at the fundamental in the
power spectrum and the presence of a peak near the second harmonic of the period.
For comparison, in IGR J16493-4348 
BAT observations also showed changes in the strength of the superorbital
modulation \citep{Coley2019}, but with the modulation preserving the phase and shape
of the modulation before and after an interval where the modulation significantly
weakened. Similarly, for 2S\,0114+650 \citet{Hu2017} reported that the superorbital 
period was stable, but that the modulation amplitude was highly variable.
In this case, \citet{Hu2017} reported that the neutron star exhibited a long
term increase in rotational frequency, but that
reductions in superorbital
modulation amplitude were associated with times when the spin frequency
was not increasing as rapidly.

For \src, measurements by \citet{Rubin1997} and \citet{Bildsten1997} with 
BATSE demonstrated that
a previous long-term spin-down trend changed to spin up, with a torque
reversal in approximately 1990.
This spin up trend was found to continue from observations made by
\citet{Baykal2006} in 2003 with the {\it Rossi X-ray Timing Explorer}.
Since then a second torque reversal occurred as shown by the GBM
pulse frequency measurements \citep{Finger2009} and also observed with
\INTEGRAL\ \citep{Hemphill2013}.
This torque reversal occurred during the BAT observations, but before the start
of the GBM observations.
The continued
observations with the GBM show that the long-term spin down initially
continued, with a more recent flattening of pulse frequency changes (Figure \ref{fig:long_lc}).
Although the long-term spin period changes of \src\ are in the opposite
direction from 2S\,0114+650, there is at least a suggestion that the flattening
of the trend might be associated with the properties of the superorbital modulation.
However, to determine whether this is more than coincidental, further similar
changes of superorbital properties that occur at the same time as changes
in \nudot\ would need to be seen.
In particular, the lack of any clear change in superorbital modulation during the
torque reversal primarily found with \INTEGRAL\ may argue against a simple connection
between superorbital modulation and \nudot.

\subsection{\igr}

For \igr, while manipulating the power spectrum by removing eclipses and adding harmonics
does show an apparently significant modulation, we do not yet consider that the superorbital
modulation is definitely present due to the amount of manipulation of the data, and because
the candidate superorbital period is not detected in the new statistically independent data
obtained beyond that in \citet{Corbet2013}. i.e. the signal is not seen in observations
obtained since MJD 56,452 (Figure \ref{fig:16393_bat_2d_power}).
{\viibf 
However, the orbital period and candidate superorbital period are consistent with the possible
correlation between these parameters (Figure \ref{fig:super_ratio}, Table \ref{table:sources}). 
}

For \src\ there is a possible connection between the change in the superorbital modulation
and the second of the two torque changes on the neutron star (Section \ref{sect:source_discuss}).
For \igr\ there are only a few measurements of the neutron star pulse period with
long intervals between them \citep{Bodaghee2016}  and the most recent measurement is from 2014-06- 27 (MJD 56,835). 
Thus, although torque changes are implied for \igr\ from these measurements, we cannot clearly associate a possible cessation
of superorbital modulation with such a change.

\subsection{Models}

\subsubsection{Model Overview}
Several mechanisms have been proposed for the origin of the superorbital modulation,
see for example the discussion in \citet{Coley2019},
including a precessing accretion disk, a triple stellar system, precession
of the donor star, corotating interaction regions (CIRs) in the stellar wind,
and tidal oscillations in the primary. While a precessing disk and a triple
system were both considered to be unlikely, the other proposed models could
not be excluded.  
The orbital and superorbital periods of \src\ are consistent with the possible
correlation between these periods previously noted \citep{Corbet2013},
although there are still only six systems with definite superorbital periods.
Such a correlation would require modulation mechanism which would produce
this, but it is still unclear how any of the proposed models would achieve this.

\subsubsection{Corotating Interaction Regions}
In the CIR model \citep{Bozzo2017} the frequency of the superorbital modulation
is the difference between the orbital frequency and the rotation frequency of
the CIR structure. 
For \src\ the superorbital period is consistent to better than 0.02\% with being
exactly four times the orbital period. 
Thus, depending on whether the CIR is rotating more or less rapidly than the orbital
frequency,
the rotation period of the CIR would
be either
2.9827 $\pm$ 0.0003 
or
4.9712 $\pm$ 0.0003 
days respectively, which are consistent with ratios of 
(4/5) and (4/3) $\times$ the 3.73 day orbital period, respectively.
However, it is
only for \src\ and not the other similar systems where such a ``resonance''
would be present. 

In the lower panel of Figure \ref{fig:cir_beat} we plot the implied rotation
periods of the CIR for each source in Table  \ref{table:sources}, including \igr, 
assuming the superorbital period is caused by the beat between the orbital and CIR periods. 
For each source two periods are mathematically possible, one longer than the orbital period
and one shorter. In Figure \ref{fig:cir_beat} very strong correlations are seen
between the CIR periods and the orbital period with linear correlation coefficients 
of 0.9999  and 0.9997  for the shorter and longer periods respectively.
While the nominal probabilities, ``$p$'', of achieving such a level of correlation with random data
are very low at \sqig10$^{-10}$ and \sqig10$^{-9}$ respectively, we note that
since the CIR frequency is derived from the orbital frequency, modified by the superorbital
frequency,  the parameters are not
completely statistically independent. 
The relationship between calculated CIR period and orbital period
can be fit with either a linear function or a power law (a linear fit
to the log of the periods):

\begin{flalign*}
P_{CIR,short} & = 0.70(1) \times P_{orb} + 0.35(3) \;\mathrm{days} \\
& = 0.87(2) \times P_{orb}^{\;0.93(1)} \\
P_{CIR,long} & = 1.72(2) \times P_{orb} - 1.3(1) \;\mathrm{days} \\
& = 1.13(4) \times P_{orb}^{\;1.14(2)} \\
\end{flalign*}

\begin{figure}
\includegraphics[width=8.0cm,angle=0]{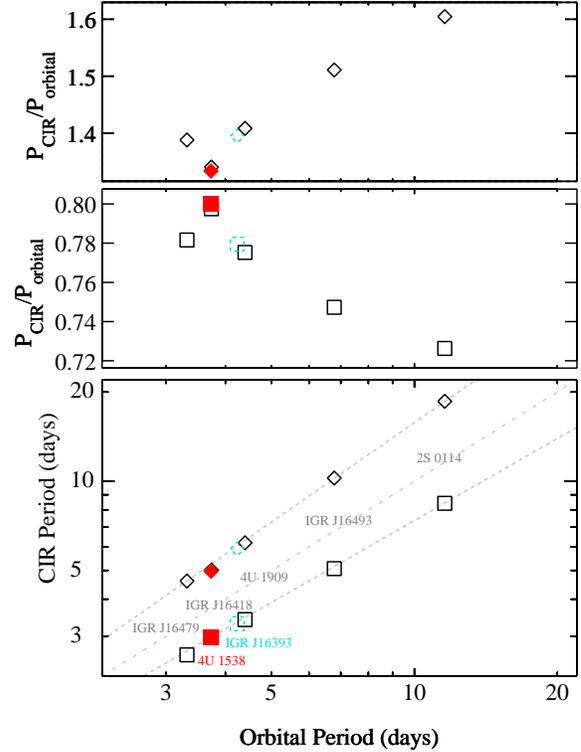}
\caption{
{\viibf
Bottom: The implied rotation period of the Corotating Interaction Region (CIR) for each
system. The boxes show the implied period if the CIR is rotating more 
slowly than the orbital period, and the diamonds are the implied period if
the CIR rotates more rapidly than the orbital period. 
The dashed lines show
linear fits in log space. The light blue dot-dashed line indicates P$_{CIR}$ = P$_{orb}$.
Note that the symbols for \src\ and IGR\,J16418$-$4532 almost completely overlap.
Middle: ratio of CIR rotation period to orbital period if the CIR is rotating more
rapidly.
Top: ratio of CIR rotation period to orbital period if the CIR is rotating more slowly.
For all panels the filled red symbols show the parameters of \src\ and the dashed cyan symbols show
the parameters of \igr.
}
}
\label{fig:cir_beat}
\end{figure}

In the middle and upper panels of  Figure \ref{fig:cir_beat} we plot the implied
CIR periods as a fraction of the orbital period for the shorter and longer CIR periods
respectively. The shorter CIR period is found to range between  
0.73 to 0.8 of the orbital period, with \src\ having the largest implied CIR period as
a fraction of orbital period. 
The longer CIR period is found to range between 1.33 and 1.60 times the orbital period,
with \src, in this case, having the lowest ratio.
For both the shorter and the longer periods there is some correlation between period
ratio and orbital period with linear correlation coefficients of $-$0.92 ($p$ = 0.003)
and 0.95 ($p$ = 0.001) respectively. \citet{Bozzo2017} pointed out that multiple CIRs might
exist in the wind. The small range of implied CIR period to orbital period that we find
suggests that, within the context of this model, while such multiple regions may cause features in the light curve,
they would not have caused us to find superorbital periods which are actually an integer fraction of
the underlying true period. 

If the CIR rotation period was identical to the rotation period of the primary star, and the
rotation of the primary was tidally locked to the orbital period, then the CIR period would be
identical to the orbital period and no beat-period superorbital modulation could be produced.
Therefore, for this model, the primary star rotation period must be non-synchronous with
the orbital period and/or the CIR rotation period must differ from the primary star's rotation
period. In addition, for all of the systems, Figure \ref{fig:cir_beat} shows that the 
relative period ratio is roughly the same for all systems.  
This CIR structure is thought to be linked to stellar spots \citep{Lobel2008}
and, at least in principle, its rotation frequency could differ from that of the primary star \citep{Bozzo2017}.

We suggest that there are two possible mechanisms where the CIR period will be close to, but not
exactly, the orbital period. First, we propose that the primary stars are at rotating at a rate
driven by tidal synchronization. However, 
for eccentric systems, tidal synchronization is driven
by the orbital speed at periastron \citep[e.g.][]{Lurie2017}.
This will result in the primary star's rotation period being ``pseudo-synchronized''
at a rate faster than for a circular orbit \citep{Hut1981,Lurie2017}.
Alternatively, the deviation of the CIR period from the orbital period
might be driven by differential rotation of the primary star.
For the isolated O4I(n)fp star $\zeta$ Pup, \citet{Ram2018} proposed
that the CIR was linked to spots at higher latitudes on the star that
were rotating {\em faster} than at lower latitudes.
We note that from Kepler observations  \citet{Lurie2017}
found a population of late-type eclipsing binaries where
stellar rotation periods were 13\% {\em slower} than synchronous, and they
attributed this to differential rotation of high-latitude star spots.
Thus, at least in those stars, differential rotation was able to persist
despite tidal synchronization of the rotation rate at lower stellar latitudes.
We cannot discriminate between the case where the CIR is rotating more rapidly
or more slowly than the orbital frequency. However, the modest spread of ratio of implied
CIR period to orbital period might be a hint that it is the same situation, i.e.
faster or slower, for all systems. 

We note that if the CIR rotation period is approximately a fixed fraction or multiple of the
orbital period, then this in itself would lead to a correlation between superorbital
and orbital periods. If the CIR period is a factor ``$F$'' times the orbital period,
then the superorbital period will be \( |(1 - 1/F)|^{-1}P_{orbital} \). 
In the case that the fractional difference between CIR period and orbital period increases
with orbital period, as is a possibility hinted at by the middle and top panels of
Figure \ref{fig:cir_beat}, then the ratio of superorbital period to orbital period
will decrease, as may be seen in the top panel of Figure \ref{fig:super_ratio}.
Systems that do not have significant co-rotating structure in their winds, or where the
CIR period is equal to the orbital period, will not exhibit superorbital modulation.
This may account for the significant number of wind-accretion HMXBs where superorbital modulation
has not yet been detected \citep{Corbet2013}.

In the CIR model, a change in the modulation properties with the second
harmonic becoming stronger would be accounted for by a change in the CIR structures
in the stellar wind, for example the appearance of multiple regions \citep{Bozzo2017}.
Such a change in superorbital modulation profiles has not yet been seen in 2S\,0114+650
or IGR J16493-4348 despite changes in the amplitude of the modulation \citep[][respectively]{Hu2017,Coley2019}.
For \igr\ the ``bumpy" shape of the light curve folded on
the candidate superorbital period (Figure \ref{fig:16393_bat_super_fold})
could perhaps be most directly explained by the CIR model with the presence of several
corotating regions in the stellar wind, each of which gives rise to peaks of different strength.

\subsubsection{Tidally Induced Pulsations}
In the tidally induced pulsation model \citep{Zahn1975,Koenigsberger2006,Moreno2005,Moreno2011}, 
this is only predicted to generate superorbital
modulation for circular systems as for eccentric systems the pulsations
would be expected to occur on the orbital period. While for \src\
the eccentricity is unclear, some of the other systems with superorbital modulation
do have detectable eccentricities (Table \ref{table:sources}).
Perhaps for \src\ the presence of pulsations of the primary in 
resonance with the orbital period
could be imagined, but this would not apply to the other systems where superorbital
and orbital periods are further from an integer ratio.

Similar to the case of the CIR model, if the rotation of the primary star is not tidally synchronized
to the orbital period, then modulation could potentially be caused by a beat between a pulsation
period and the orbital period. In this case, the ``CIR'' periods in Figure \ref{fig:cir_beat}
could be interpreted as intrinsic pulsation periods.

\subsubsection{General Considerations}
While there also appear to be correlations between superorbital and
orbital period for HMXBs powered by Roche-lobe overflow, and for
Be star systems \citep{Corbet2013, Townsend2020} these other systems are
likely to be physically different as we do not expect the presence
of a persistent accretion disk in wind-accreting HMXBs, although for some
such it may be possible that a transient accretion disk could form \citep[e.g.][]{Xu2019}.

There are also other clues that hint at the nature of the mechanism.
Similar to what has been seen in IGR J16493-4348 \citep{Coley2019} and
2S\,0114+650 \citep{Hu2017}, the strength of the superorbital modulation
is variable, and the most recent observations with the BAT show
a change in the superorbital modulation profile. At the same time, pulse frequency measurements
with the \Fermi\ GBM show that the spin-down rate of the pulsar decreased.
If this is a persistent feature of the system then it would be similar to
that noted by \citet{Hu2017} for 2S\,0114+650 who reported that when the system
was no longer spinning {\em} up, the superorbital modulation amplitude
decreased. 
However, we note that the large spin-up event that occurred during the initial part of
the BAT observations was not accompanied by any apparent change in superorbital modulation.
\citet{Hu2017} attributed the apparent connection between superorbital modulation
and spin period changes to the behavior of a transient
accretion disk in 2S\,0114+650, and this is also discussed by \citet{Wang2020}.
The presence of such a disk in \src\ seems
less likely as no evidence for this has been seen from optical spectroscopy \citep{Reynolds1992}. 
For IGR J16493-4348, when the modulation amplitude decreased,
and then increased again, the phasing of the modulation was unchanged, implying
that the system had a ``memory'' of this, which is challenging to explain 
with a transient accretion disk. 
Another long-term change that has been observed in \src\ is the energy of
a cyclotron resonance scattering feature in its X-ray spectrum, although
\citet{Hemphill2019} consider that it is not yet possible to definitely associate
changes in this with torque reversals on the neutron star.
Even if \src\ does not possess a, perhaps transient, accretion disk, there do nevertheless
appear to be changes in the nature of the accretion process that result in spin-down
and spin-up trends that continue for durations of \sqig years.

The optical photometry of \src/QV\,Nor, similarly to previous observations, shows the presence
of significant ellipsoidal modulation due to the tidal distortion of the primary
star by the neutron star. While there are no explicit predictions of the degree of
optical modulation on the superorbital period for any model, some models may have at least
the possibility of resulting in some level of modulation. 
For example, a transient disk may be expected to produce optical emission,
and tidally induced oscillations should also produce changes in optical brightness
at some level. 
While our optical photometry does not show any significant modulation at 
the superorbital period, the observations were obtained after the 14.9130 day 
modulation in the BAT light curve was no longer strongly present.
Continued optical monitoring of \src\ would both enable a steady increase
in the sensitivity to detection of periodic optical modulation on the superorbital
period and would be important to compare if/when the 14.9130 day X-ray modulation
strengthens again.

\section{Conclusion}

\src\ is now an additional member of the class of wind-accretion supergiant HMXBs
which show superorbital modulation. The orbital and superorbital periods are
consistent with the possible correlation between these that we previously
noted. The driving mechanism for superorbital modulation remains
mysterious and such a correlation is challenging to explain.
However, a model based on the beat between a CIR and the orbit,
where the CIR period has a modest deviation from tidal synchronization due to
either orbital eccentricity or differential rotation of the primary, appears
promising. 

If \src\ was fainter, it would not have been possible to detect
the superorbital modulation as the superorbital peak in the power spectrum is much smaller than
the orbital peak. 
The difficulty of detecting superorbital modulation is compounded
when the modulation can either decrease in strength, as in IGR J16493-4348 \citep{Coley2019} or
2S\,0114+650 \citep{Hu2017}, or change modulation shape, as for \src.
Thus the possibility remains that low-level superorbital
modulation could be present in other similar systems where it has
not yet been seen because they are either fainter and/or the superorbital modulation
is less persistent.

For \igr, although it is not yet certain that the candidate modulation is real,
it would be similar to the other superorbital
systems in that 
it would also be variable in its properties. In addition, its superorbital and orbital
periods would be consistent with the possible correlation between these periods (Figure \ref{fig:super_ratio}).
If the modulation is real, and the physical conditions that cause it,
whether it is the presence of CIRs, a transient accretion disk, or something else,
return we would expect to see the superorbital modulation again. In that case, an investigation
of other properties of the system, including its X-ray spectrum and any accretion torque changes, may be
revealing.

Continued long-term observations of both \src\ and \igr\ have the potential to give insights
into the physics of superorbital modulation in supergiant HMXBs.
The relative brightness of \src\ in X-rays is advantageous for continued long-term studies.
If we can continue to monitor 
both the superorbital modulation and the pulse frequency
changes to further explore a tentative connection between these.
If \src\ is about to enter an extended spin-up phase we will have the opportunity
to investigate the effects of this on the superorbital modulation.
The brightness of the optical counterpart QV Nor also facilitates long-term
photometric monitoring and so a search for any long-term changes in this waveband.

}

\acknowledgements
This work makes use of optical observations from the LCOGT network, and we thank the XB-NEWS team
and especially Dan Bramich and Dave Russell for
providing the photometric analysis of these and advice. This paper made use of Swift/BAT transient
monitor results provided by the Swift/BAT team and Fermi GBM results provided by the Fermi GBM team.
We thank Malcolm Coe for productive discussion on optical monitoring and Nazma Islam for useful
discussion on superorbital models.
We also thank the referee for valuable comments.

\pagebreak

\begin{turnpage}

\begin{deluxetable}{lccccccc}
%\rotate
\tablewidth{7in}

\tablecaption{LCO Optical Observations of QV Nor/\src}
\tablehead{
\colhead{Site} & \colhead{Telescope} & \colhead{Instrument} &
\colhead{Filter} & \colhead{Start Night} & \colhead{End Night} & \colhead{Number} & \colhead{Exposure (s)}\\
}
\startdata
SAAO & 1m010 &  fl16 &  B &     2018-08-15 &    2018-09-22 &     12 & 15\\
SAAO & 1m010 &  fl16 &  V &     2018-08-15 &    2018-09-22 &     1 & 5\\
SAAO & 1m012 &  fl06 &  B &     2018-08-22 &    2018-09-28 &     5 & 15\\
SAAO & 1m012 &  fl06 &  V &     2018-08-22 &    2018-09-28 &     5 & 5\\
SSO & 1m003 &   fl11 &  B &     2018-09-21 &    2018-09-29 &     5 & 15\\
SSO & 1m003 &   fl11 &  V &     2018-09-21 &    2018-09-29 &     4 & 5\\
SSO & 1m011 &   fl12 &  B &     2018-08-20 &    2018-08-21 &     2 & 15\\
SSO &   1m011 & fl12 &  V &     2018-08-20 &    2018-08-21 &     2 & 5\\
\hline
SSO &   2m002 & fs01 &  B &     2019-05-21 &    2019-08-29 &     63 & 30\\
SSO &   2m002 & fs01 &  V &     2019-05-21 &    2019-08-29 &     63 & 15\\
\enddata

\tablecomments{Site: SAAO = South African Astronomical Observatory, SSO = Siding Spring Observatory.\\
Telescope: all ``1m'' are 1-m, and all ``2m'' are 2-m aperture.\\
Instrument: all ``fl'' are Sinistro imagers and all ``fs'' are Spectral imagers.\\
For details see \citet{Brown2013} and \citet{Russell2019}. 
}
\label{table:lco_observations}
\end{deluxetable}

\clearpage

\begin{deluxetable}{lccccccc}
%\rotate
\tablewidth{8.7in}

\tablecaption{Wind-Accretion Supergiant HMXBs with Periodic Superorbital Modulation}

\tablehead{
\colhead{Name} & \colhead{P$_{orb}$} & \colhead{P$_{super}$} &
\colhead{P$_{super}$/P$_{orb}$} &
\colhead{P$_{spin}$} & \colhead{Spectral Type} & \colhead{Eccentricity} &\colhead{SFXT?}\\
\colhead{} & \colhead{(days)} & \colhead{(days)} &
\colhead{} &
\colhead{(s)} & \colhead{} & \colhead{} \\
}

\startdata
IGR J16479-4514 &  3.31961 $\pm$ 0.00004 (a) &  11.880 $\pm$ 0.002 & 3.58 & ? & O8.5 I/O9.5 Iab & ? & Y \\
{\bf 4U 1538-52} & {\bf 3.728354 $\pm$ 0.000009} & {\bf 14.9130 $\pm$ 0.0026} & {\bf 4.00} & {\bf 526} & {\bf B0 Iab} & {\bf $<$ 0.2} & {\bf N}  \\ 
IGR J16418-4532 &  3.73881 $\pm$ 0.00002 (a) &  14.730 $\pm$ 0.006 &  3.94 & 1212 & {\mybf O8.5} & ? & N (b) \\
4U 1909+07 & 4.4003 $\pm$ 0.0004  &  15.180 $\pm$ 0.003 & 3.45 & 605 & B0-3 I (c) & 0.02 $\pm$ 0.04 & N \\
IGR J16493-4348  &  6.7828 $\pm$ 0.0004 (d) & 20.058 $\pm$ 0.007 (e) & 2.96 & 1093 & B0.5 Ia (d) & $<$0.2 (d) & N \\
2S 0114+650  &  11.591  $\pm$ 0.003 & 30.76 $\pm$ 0.03 & 2.65  & \sqig9700 & B1 Ia & 0.18 $\pm$ 0.05 & N\\
\tableline
IGR J16393-4643 & 4.23810 $\pm$ 0.00007 (a) & (14.981 $\pm$ 0.002) & (3.53) &  910 & ? & ? & N \\
\enddata
\tablecomments{The superorbital period for IGR J16393-4643 is considered to be a candidate
and not a definite detection. The table is updated from Table 1 in \citet{Corbet2013} based on: 
(a) \citet{Coley2015},
(b) \citet{Romano2015},
(c) \citet{Martinez2015},
(d) \citet{Pearlman2019},
(e) \citet{Coley2019}.
For parameters of \src, apart from the superorbital period, see Section \ref{sect:source_intro}.
The SFXT column indicates whether a source is known to be a Supergiant Fast X-ray Transient.
}
\label{table:sources}
\end{deluxetable}
\end{turnpage}


\begin{thebibliography}{}

% The Burst Alert Telescope (BAT) on the SWIFT Midex Mission
\bibitem[Barthelmy et al.(2005)]{Barthelmy2005} Barthelmy, S.~D., 
Barbier, L.~M., Cummings, J.~R., et al.\ 2005, \ssr, 120, 143

% Recent timing studies on RXTE observations of 4U 1538-52 
\bibitem[Baykal et al.(2006)]{Baykal2006} 
Baykal, A., Inam, S. {\c{C}}., \& Beklen, E.\ 2006, \aap, 453, 1037

% Observations of Accreting Pulsars 
\bibitem[Bildsten et al.(1997)]{Bildsten1997} 
Bildsten, L., Chakrabarty, D., Chiu, J., et al.\ 1997, \apjs, 113, 367

%  Chandra Observations of Five INTEGRAL Sources: New X-Ray Positions for IGR J16393-4643 and IGR J17091-3624 
\bibitem[Bodaghee et al.(2012)]{Bodaghee2012} 
Bodaghee, A., Rahoui, F., Tomsick, J.~A., et al.\ 2012, \apj, 751, 113

{\viibf
% NuSTAR Discovery of a Cyclotron Line in the Accreting X-Ray Pulsar IGR J16393-4643 
\bibitem[Bodaghee et al.(2016)]{Bodaghee2016} Bodaghee, A., Tomsick, J.~A., Fornasini, F.~M., et al.\ 2016, \apj, 823, 146
}

%  The super-orbital modulation of supergiant high-mass X-ray binaries 
\bibitem[Bozzo et al.(2017)]{Bozzo2017} 
Bozzo, E., Oskinova, L., Lobel, A., et al.\ 2017, \aap, 606, L10

% Las Cumbres Observatory Global Telescope Network 
\bibitem[Brown et al.(2013)]{Brown2013} 
Brown, T.~M., Baliber, N., Bianco, F.~B., et al.\ 2013, \pasp, 125, 1031

% The Orbit of the Binary X-Ray Pulsar 4U 1538-52 from Rossi X-Ray Timing Explorer Observations 
\bibitem[Clark(2000)]{Clark2000} Clark, G.~W.\ 2000, \apjl, 542, L131

{\viibf
% semi-weighted mean 1
\bibitem[Cochran (1937)]{Cochran1937} Cochran, W.~G.\ 1937, Supplement to the Journal of the Royal Statistical Society,
4, 102
}

{\viibf
% semi-weighted mean 2
\bibitem[Cochran (1954)]{Cochran1954} Cochran, W.~G.\ 1954, Biometrics, 10, 101
}

%  Probing the Masses and Radii of Donor Stars in Eclipsing X-Ray Binaries with the Swift Burst Alert Telescope 
\bibitem[Coley et al.(2015)]{Coley2015} 
Coley, J.~B., Corbet, R.~H.~D., \& Krimm, H.~A.\ 2015, \apj, 808, 140

% A Study of the 20 Day Superorbital Modulation in the High-Mass X-ray Binary IGR J16493-4348 
\bibitem[Coley et al.(2019)]{Coley2019} 
Coley, J.~B., Corbet, R.~H.~D., F{\"u}rst, F., et al.\ 2019, \apj, 879, 34

% The orbit and pulse period of X 1538-522 from GINGA observations. 
\bibitem[Corbet et al.(1993)]{Corbet1993} 
Corbet, R.~H.~D., Woo, J.~W., \& Nagase, F.\ 1993, \aap, 276, 52

% A 6.8 Day Period in IGR J16493-4348 from Swift/BAT and RXTE/PCA Observations
\bibitem[Corbet et al. (2010)]{Corbet2010} Corbet, R.~H.~D., 
Barthelmy, S.~D., Baumgartner, W.~H., Krimm, H.~A., Markwardt, C.~B., 
Skinner, G.~K., \& Tueller, J.\ 2010, The Astronomer's Telegram, 2599, 1 

% Superorbital Periodic Modulation in Wind-accretion High-mass X-Ray Binaries from Swift Burst Alert Telescope Observations
\bibitem[Corbet \& Krimm(2013)]{Corbet2013} 
Corbet, R.~H.~D., \& Krimm, H.~A.\ 2013, \apj, 778, 45

% Diverse Long-term Variability of Five Candidate High-mass X-Ray Binaries 
%from Swift Burst Alert Telescope Observations 
\bibitem[Corbet et al.(2017)]{Corbet2017} 
Corbet, R.~H.~D., Coley, J.~B., \& Krimm, H.~A.\ 2017, \apj, 846, 161

% Apparent Superorbital Modulation in 4U 1538-52 at Four Times the Orbital Period 
%\bibitem[Corbet et al.(2018)]{Corbet2018} 
%Corbet, R.~H.~D., Coley, J.~B., Krimm, H.~A., et al.\ 2018, The Astronomer's Telegram, 11918, 1

% Super-orbital period in the high-mass X-ray binary 2S 0114+650
\bibitem[Farrell et al.(2006)]{Farrell2006} Farrell, S.~A., Sood, 
R.~K., \& O'Neill, P.~M.\ 2006, \mnras, 367, 1457 

% A detailed study of 2S 0114+650 with the Rossi X-ray Timing Explorer
\bibitem[Farrell et al.(2008)]{Farrell2008} Farrell, S.~A., Sood, 
R.~K., O'Neill, P.~M., \& Dieters, S.\ 2008, \mnras, 389, 608 

%  Long-term Monitoring of Accreting Pulsars with Fermi GBM 
\bibitem[Finger et al.(2009)]{Finger2009} 
Finger, M.~H., Beklen, E., Narayana Bhat, P., et al.\ 2009, arXiv e-prints, arXiv:0912.3847

%  Measurements of Cyclotron Features and Pulse Periods in the 
% High-mass X-Ray Binaries 4U 1538-522 and 4U 1907+09 with the 
% International Gamma-Ray Astrophysics Laboratory 
\bibitem[Hemphill et al.(2013)]{Hemphill2013} 
Hemphill, P.~B., Rothschild, R.~E., Caballero, I., et al.\ 2013, \apj, 777, 61

% The First NuSTAR Observation of 4U 1538-522: Updated Orbital Ephemeris and a Strengthened Case for an Evolving Cyclotron Line Energy 
\bibitem[Hemphill et al.(2019)]{Hemphill2019}
Hemphill, P.~B., Rothschild, R.~E., Cheatham, D.~M., et al.\ 2019, \apj, 873, 62

% A prescription for period analysis of unevenly sampled time series
\bibitem[Horne 
\& Baliunas(1986)]{Horne1986} Horne, J.~H., \& Baliunas, S.~L.\ 1986, \apj, 302, 757

% Evolution of Spin, Orbital, and Superorbital Modulations of 4U 0114+650 
\bibitem[Hu et al.(2017)]{Hu2017} 
Hu, C.-P., Chou, Y., Ng, C.-Y., et al.\ 2017, \apj, 844, 16

{\viibf
%  Tidal evolution in close binary systems. 
\bibitem[Hut(1981)]{Hut1981} Hut, P.\ 1981, \aap, 99, 126
}


% Optical light curve for the X-ray binary 4U 1538-52
\bibitem[Ilovaisky et al.(1979)]{Ilovaisky1979} 
Ilovaisky, S.~A., Chevalier, C., \& Motch, C.\ 1979, \aap, 71, L17

% Orbital Decay and Evidence of Disk Formation in the X-Ray Binary Pulsar OAO 1657-415
\bibitem[Jenke et al.(2012)]{Jenke2012} 
Jenke, P.~A., Finger, M.~H., Wilson-Hodge, C.~A., et al.\ 2012, \apj, 759, 124

% Investigating a unique partial eclipse in the high-mass 
% X-ray binary IGR J16393-4643 with Swift-XRT 
\bibitem[Kabiraj et al.(2020)]{Kabiraj2020} 
Kabiraj, S., Islam, N., \& Paul, B.\ 2020, \mnras, 491, 1491

%  Significance Testing of Periodogram Ordinates 
\bibitem[Koen(1990)]{Koen1990} Koen, C.\ 1990, \apj, 348, 700

% The X-ray binary 2S0114+650=LSI+65 010. 
% A slow pulsar or tidally-induced pulsations?
\bibitem[Koenigsberger et al.(2006)]{Koenigsberger2006} 
Koenigsberger, G., Georgiev, L., Moreno, E., et al.\ 2006, \aap, 458, 513

% Characterizing X-ray binary long-term variability
\bibitem[Kotze 
\& Charles(2012)]{Kotze2012} Kotze, M.~M., \& Charles, P.~A.\ 2012, \mnras, 420, 1575 

% The Swift/BAT Hard X-ray Transient Monitor
\bibitem[Krimm et al.(2013)]{Krimm2013}
Krimm, H.~A., Holland, 
S.~T., Corbet, R.~H.~D., et al.\ 2013, \apjs, 209, 14 

{\viibf
%  Modeling Ultraviolet Wind Line Variability in Massive Hot Stars 
\bibitem[Lobel \& Blomme(2008)]{Lobel2008} Lobel, A. \& Blomme, R.\ 2008, \apj, 678, 408
}

%  Tidal Synchronization and Differential Rotation of Kepler Eclipsing Binaries 
\bibitem[Lurie et al.(2017)]{Lurie2017} 
Lurie, J.~C., Vyhmeister, K., Hawley, S.~L., et al.\ 2017, \aj, 154, 250

% Spectra and Pulse Period of the Binary X-Ray Pulsar 4U 1538-52
\bibitem[Makishima et al.(1987)]{Makishima1987} 
Makishima, K., Koyama, K., Hayakawa, S., et al.\ 1987, \apj, 314, 619

% The Ups & Downs of Accreting X-ray Pulsars: Decade-long observations with the Fermi Gamma-ray Burst Monitor 
\bibitem[Malacaria et al.(2020)]{Malacaria2020} 
Malacaria, C., Jenke, P., Roberts, O.~J., et al.\ 2020, arXiv e-prints, arXiv:2004.00051

%  The donor star of the X-ray pulsar X1908+075
\bibitem[Mart{\'\i}nez-N{\'u}{\~n}ez et al.(2015)]{Martinez2015} 
Mart{\'\i}nez-N{\'u}{\~n}ez, S., Sander, A., G{\'\i}menez-Garc{\'\i}a, A., et al.\ 2015, \aap, 578, A107


% The Fermi Gamma-ray Burst Monitor 
\bibitem[Meegan et al.(2009)]{Meegan2009} 
Meegan, C., Lichti, G., Bhat, P.~N., et al.\ 2009, \apj, 702, 791

%  Line profile variability from tidal interactions in binary systems 
\bibitem[Moreno et al.(2005)]{Moreno2005} 
Moreno, E., Koenigsberger, G., \& Toledano, O.\ 2005, \aap, 437, 641

%  Eccentric binaries. Tidal flows and periastron events 
\bibitem[Moreno et al.(2011)]{Moreno2011} 
Moreno, E., Koenigsberger, G., \& Harrington, D.~M.\ 2011, \aap, 528, A48

% Orbital Evolution and Orbital Phase Resolved Spectroscopy of the HMXB Pulsar 4U 1538-52 with RXTE-PCA and BeppoSAX 
\bibitem[Mukherjee et al.(2006)]{Mukherjee2006} 
Mukherjee, U., Raichur, H., Paul, B., et al.\ 2006, Journal of Astrophysics and Astronomy, 27, 411

% Optical photometry of massive X-ray binaries : 4U 1538-52/QV Nor. 
\bibitem[Pakull et al.(1983)]{Pakull1983} 
Pakull, M., van Amerongen, S., Bakker, R., et al.\ 1983, \aap, 122, 79

% The Orbital Parameters of the Eclipsing High-mass X-Ray Binary Pulsar IGR J16493-4348 from Pulsar Timing 
\bibitem[Pearlman et al.(2019)]{Pearlman2019} 
Pearlman, A.~B., Coley, J.~B., Corbet, R.~H.~D., et al.\ 2019, \apj, 873, 86

{\viibf
%  BRITE-Constellation high-precision time-dependent photometry of the early O-type supergiant ζ Puppis unveils the photospheric drivers of its small- and large-scale wind structures 
\bibitem[Ramiaramanantsoa et al.(2018)]{Ram2018} Ramiaramanantsoa, T., Moffat, A.~F.~J., Harmon, R., et al.\ 2018, \mnras, 473, 5532
}

% Refined Neutron Star Mass Determinations for Six Eclipsing X-Ray Pulsar Binaries 
\bibitem[Rawls et al.(2011)]{Rawls2011} 
Rawls, M.~L., Orosz, J.~A., McClintock, J.~E., et al.\ 2011, \apj, 730, 25

% Optical spectroscopy of the massive X-ray binary QV Nor (4U 1538-52). 
\bibitem[Reynolds et al.(1992)]{Reynolds1992}
Reynolds, A.~P., Bell, S.~A., \& Hilditch, R.~W.\ 1992, \mnras, 256, 631

%  Seven years with the Swift Supergiant Fast X-ray Transients project 
\bibitem[Romano(2015)]{Romano2015} 
Romano, P.\ 2015, Journal of High Energy Astrophysics, 7, 126

%  Giant outburst from the supergiant fast X-ray transient IGR J17544-2619: accretion from a transient disc?
\bibitem[Romano et al.(2015)]{Romano2015b} 
Romano, P., Bozzo, E., Mangano, V., et al.\ 2015, \aap, 576, L4

% Observation of a Long-Term Spin-up Trend in 4U 1538-52 
\bibitem[Rubin et al.(1997)]{Rubin1997} 
Rubin, B.~C., Finger, M.~H., Scott, D.~M., et al.\ 1997, \apj, 488, 413

%  Optical precursors to X-ray binary outbursts (XB-NEWS)
\bibitem[Russell et al.(2019)]{Russell2019} 
Russell, D.~M., Bramich, D.~M., Lewis, F., et al.\ 2019, Astronomische Nachrichten, 340, 278

% Studies in astronomical time series analysis. II - Statistical aspects of spectral analysis of unevenly spaced data
\bibitem[Scargle(1982)]{Scargle1982} Scargle, J.~D.\ 1982, \apj, 
263, 835 

{\viibf
% Studies in astronomical time series analysis. III - Fourier transforms, autocorrelation functions, 
% and cross-correlation functions of unevenly spaced data
\bibitem[Scargle(1989)]{Scargle1989} Scargle, J.~D.\ 1989, \apj, 
343, 874 
}

% On Nonsteady Accretion in Stellar Wind--fed X-Ray Sources 
\bibitem[Taam \& Fryxell(1988)]{Taam1988} Taam, R.~E., \& Fryxell, B.~A.\ 1988, \apjl, 327, L73

%  Numerical Studies of Asymmetric Adiabatic Accretion Flow: The Effect of Velocity Gradients 
\bibitem[Taam \& Fryxell(1989)]{Taam1989} Taam, R.~E., \& Fryxell, B.~A.\ 1989, \apj, 339, 297

%  Orbital and superorbital periods in ULX pulsars, disc-fed HMXBs, Be/X-ray binaries, and double-periodic variables 
\bibitem[Townsend \& Charles(2020)]{Townsend2020} Townsend, L.~J., \& Charles, P.~A.\ 2020, arXiv e-prints, arXiv:2004.14207

%  Understanding the coexistence of spin-up and spin-down behaviours in long-period X-ray pulsars 
\bibitem[Wang \& Tong(2020)]{Wang2020} Wang, W., \& Tong, H.\ 2020, \mnras, 492, 762

%  Bondi-Hoyle-Lyttleton accretion in supergiant X-ray binaries: stability and disc formation
\bibitem[Xu \& Stone(2019)]{Xu2019} 
Xu, W., \& Stone, J.~M.\ 2019, \mnras, 488, 5162

% The dynamical tide in close binaries.
\bibitem[Zahn(1975)]{Zahn1975} 
Zahn, J.-P.\ 1975, \aap, 41, 329 


\end{thebibliography}
\end{document}